\documentclass[aps,prd,preprintnumbers]{revtex4}
\usepackage[english]{babel}
\usepackage{amssymb}
\usepackage{amsfonts}
\usepackage{amsmath}
\usepackage{graphicx}
\usepackage{epsfig}
\usepackage{mathrsfs}
\usepackage{verbatim}

\newcommand{\be}{\begin{equation}} \newcommand{\ee}{\end{equation}}
\newcommand{\bea}{\begin{eqnarray}} \newcommand{\eea}{\end{eqnarray}}
\newcommand{\beann}{\begin{eqnarray*}}  \newcommand{\eeann}{\end{eqnarray*}}
\newcommand{\bfig}{\begin{figure}} \newcommand{\efig}{\end{figure}}
\newcommand{\ba}{\begin{array}} \newcommand{\ea}{\end{array}}
\newcommand{\bcen}{\begin{center}} \newcommand{\ecen}{\end{center}}
\newcommand{\btab}{\begin{tabular}} \newcommand{\etab}{\end{tabular}}

\def\tr{\operatorname{tr\:}}

\newcommand{\bra}[1]{\langle #1|}
\newcommand{\ket}[1]{|#1\rangle}

\newtheorem{Proposition}{Proposition}[section]

\newtheorem{Theorem}{Theorem}[section]
\newtheorem{Lemma}{Lemma}[section]
\newtheorem{Corrolary}{Corrolary}[section]

\newcommand{\bp}{\begin{Proposition}}   \newcommand{\ep}{\end{Proposition}}
\newcommand{\bt}{\begin{Theorem}}   \newcommand{\et}{\end{Theorem}}
\newcommand{\bl}{\begin{Lemma}}     \newcommand{\el}{\end{Lemma}}
\newcommand{\bc}{\begin{Corrolary}} \newcommand{\ec}{\end{Corrolary}}

\def\az{{{A^{(0)}}}}
\def\ao{{{A^{(1)}}}}
\def\at{{{A^{(2)}}}}

\def\bz{{{B^{(0)}}}}
\def\bo{{{B^{(1)}}}}
\def\bt{{{B^{(1)}}}}

\def\Tz{{{T^{(0)}}}}
\def\To{{{T^{(1)}}}}
\def\Tt{{{T^{(2)}}}}

\def\tTz{{{\overline{T}}^{(0)}}}
\def\tTo{{{\overline{T}}^{(1)}}}
\def\tTt{{{\overline{T}}^{(2)}}}

\newcommand\tlpm{\tilde{\lambda}_\pm}

\newcommand\tlp{\frac{\lambda}{8 g_B}}

\begin{document}
\title{\Large An improved model of vector mesons in holographic QCD}


\author{ Raul Alvares}
\email{raulalvares@gmail.com}
\affiliation{ School of Physics and Astronomy, ~\,Southampton University, ~\,Southampton,~\,SO17 1BJ, ~\,UK}
\author{Carlos Hoyos}
\email{choyos@phys.washington.edu} 
\author{Andreas Karch}
\email{karch@phys.washington.edu}
\affiliation{Department of Physics, University of Washington, ~\,Seattle, WA 98915-1560, USA}
\preprint{SHEP-11-20}

\begin{abstract}
We analyze the sector of dimension-three vector meson operators in the ``hard wall'' model of holographic QCD, including the vector and axial currents, dual to gauge fields in the bulk, and the tensor operator $\overline{\psi} \sigma^{\mu\nu}\psi$, dual to a two-form field satisfying a complex self-duality condition. The model includes the effect of chiral symmetry breaking on vector mesons, that involves a coupling between the dual gauge field and the two-form field. We compute the leading logarithmic terms in the operator product expansion of two-point functions and the leading non-perturbative contribution to the tensor-vector correlator. The result is consistent with the operator product expansion of QCD. We also study the spectrum of vector mesons numerically.
\end{abstract}

\maketitle


\section{Introduction}


Popular phenomenological models of QCD, such as the ``hard wall model" \cite{Erlich:2005qh,DaRold:2005zs} as well as ``the soft wall model" \cite{Karch:2006pv}, rely on the assumption that one can map QCD to an effective theory in the bulk of the holographic fifth dimension. This is a very strong assumption, which is not fully justified. Such an effective description implies a large hierarchy of scales between meson masses and flux tube tension, which is not present in QCD. This is the main reason why these are at best phenomenological models. Nevertheless, they are a useful resource as they provide quick and easy ways to estimate many quantitative properties, provided one is willing to live with errors which, in most instances where these models can be compared to data,
turn out to be in the 10--30\% range. These simple tools are nice to have for QCD. They are even more valuable when studying QCD-like theories in particle physics, most notably as a theory of technicolor, or more generally as a potential ``hidden sector" which may leave an imprint on LHC data. As far as, for example, meson spectra in QCD are concerned, holographic models will never be competitive with lattice gauge theories. However, when exploring theories of technicolor or hidden sectors one does not know, a priori, what the correct Lagrangian is. So one needs to explore many different models, each of which would require years of extensive computer simulations on the lattice, but only days in a holographic model.

While using a 5D effective theory for QCD is not necessarily justified, it should at least be done self-consistently. In the 5D effective theory, not only are higher derivative interactions and high dimension operators suppressed --- one is also suppressing an infinite number of additional fields. In holography, we know that every boundary operator should correspond to a field in the bulk. Fields kept in the simplest holographic models correspond to boundary operators of UV dimension\footnote{In an asymptotically free gauge theory such as QCD the dimensions of operators evolve with scale together with the coupling constant. In the UV the theory becomes free and all operators take on the free field dimensions.} 3. One can say that bulk fields dual to operators of higher dimension,
which are more massive and hence can be integrated out, are being neglected. In a top-down holographic theory, such as the AdS$_5 \times S^5$ dual of large ${\cal N}=4$ SYM with a large number of colors $N_c$ and at strong 't Hooft coupling $\lambda$ \cite{Maldacena:1997re,Gubser:1998bc,Witten:1998qj}, a similar reduction to a small subset of boundary operators and hence bulk fields is entirely justified: there is only a finite number of ten-dimensional fields dual to BPS operators, which retain their free field dimension. All other operators acquire anomalous dimensions of order $\lambda^{1/4}$, so their dual fields in the bulk acquire masses of the same order in $\lambda$ and can safely be integrated out. In the phenomenological approach one includes the fields dual to dimension 3 operators, but hopes to be able to neglect operators of dimensions 4, 5, 6 and so on.

There are, however, two additional dimension 3 operators, $\bar{\psi} \sigma^{\mu\nu} \psi$
and $\bar{\psi} \sigma^{\mu\nu} \gamma_5 \psi$ ($\sigma^{\mu\nu}=i/2[\gamma^\mu,\gamma^\nu]$ being the antisymmetrized product of two gamma matrices), whose dual field is not included in the simplest bulk model. The corresponding field, a complex bifundamental anti-symmetric rank-two tensor field $B_{\mu \nu}$, should be included for self-consistency of the model. Real and imaginary part of this field correspond to the tensor operators with and without $\gamma_5$ insertion respectively. An immediate benefit resulting from the inclusion of this extra field is that one will obtain masses for isospin triplet vector mesons with $J^{PC}=1^{+-}$, starting with the $b_1$ at a mass of 1235 MeV. They clearly should be part of the setup which, as it stands, can otherwise only incorporate $1^{--}$ and $1^{++}$ vector mesons like the $\rho$ and $a_1$.

Including a new field in the Lagrangian comes with new interaction terms and coupling constants. The original work on the hard wall model in ref. \cite{Erlich:2005qh} has proposed a rigorous procedure to fix those, which so far has been very successful phenomenologically: first of all, we assume that the bulk is described by an effective field theory so we only write down interaction terms of bulk field theory dimension 5 or less (not to be confused with the dimension of boundary operators, which after all maps to the mass of the bulk field). The corresponding coupling constants as well as the normalization of the kinetic terms and the masses are obtained by demanding that at large momentum correlation functions in the bulk agree with the corresponding field theory values at weak coupling. This is basically the statement that one trusts the effective field theory picture in the bulk all the way into the UV. While this is a very strong assumption, it at least is a hypothesis that can be tested as it allows one to make predictions for particle masses and decay constants based on very few inputs. In the original hard wall model this procedure was used to fix the mass of the vectors and the scalar in the bulk, the five-dimensional gauge coupling and, as was pointed out later in ref. \cite{Cherman:2008eh}, it was also needed to fix the map between the asymptotic form of the scalar field and the quark mass, as well as the quark condensate $\langle\bar{\psi} \psi\rangle$ in the field theory. Out of these, only the mass of the vectors could have been justified\footnote{A gauge field always has to be massless; the dual field theory current is conserved and so has dimension 3 protected by a Ward identity} without the assumption of a valid effective theory in the UV. But already for the determination of the gauge coupling one had to rely on the two-point functions of the currents, whose overall normalization is not protected.

Preliminary advances in including the $B_{\mu\nu}$ field in the bulk has recently appeared in the pioneering work \cite{Cappiello:2010tu}, but only in a form which does not account for chiral symmetry breaking. In that work only the real part of $B_{\mu \nu}$ is considered; this field propagates the new $1^{+-}$ tensor mesons, but also an additional copy of $1^{--}$ vector mesons. More importantly, only quadratic terms were included in the action for the new $B$-field. Its mass is fixed to 1 in AdS units by requiring that the dual operator has dimension 3; the normalization of the kinetic term is fixed, as in the original hard wall model, by requiring the large momentum limit of two-point functions to agree with asymptotically free QCD. However there is one more bulk operator of dimension 5 or less that needs to be included in the action to communicate the effects of chiral symmetry breaking to the $B_{\mu \nu}$ field and ensure that there is only a single set of vector mesons and no double counting, that would happen since gauge fields and the $B_{\mu \nu}$ fields have the same degenerate spectra in the absence of this term. This extra term has the form $Tr(X^\dagger F_L B + B F_R X^\dagger+h.c.)$, where
$X$ is the bifundamental scalar responsible for chiral symmetry breaking and $F_{L/R}$ are the field strengths for the bulk gauge fields dual to the chiral symmetry currents. The coupling constant in front of this term may be fixed by demanding the correct OPE structure of the correlator in the UV.

More progress in this direction was made in ref. \cite{Domokos:2011dn}, where it was proposed that the action of the complex $B_{\mu\nu}$ field should be first order, in such a way that the four-dimensional components satisfy a complex self-duality condition. The reason behind this choice is that in four dimensions the tensor operators $\bar{\psi} \sigma^{\mu\nu} \psi$ and $\bar{\psi} \sigma^{\mu\nu} \gamma_5 \psi$ are not independent, but given the definition of $\gamma_5=i\gamma^0\gamma^1\gamma^2\gamma^3$, they are related by
\begin{equation}\label{eq:dualitycond}
\bar{\psi} \sigma^{\mu\nu} \gamma_5 \psi  = \frac{i}{2}\epsilon^{\mu\nu}_{\ \ \alpha\beta} \bar{\psi} \sigma^{\alpha\beta} \psi.
\end{equation}
Hence a similar condition must be imposed on the two-form field. Following a similar procedure, we choose the action for the $B_{\mu \nu}$ to be a Chern-Simons action with a mass term,\footnote{We could also add a kinetic term of the form $(dB)^2$ to the action, we will comment more on this in section \ref{ss:equations}.} schematically
\begin{equation}
S_B = -\int d^5 x\left[ \, i(B\wedge dB^\dagger-B^\dagger \wedge dB)+m_B |B|^2\right].
\end{equation}
The duality condition follows from the equations of motion. We differ from ref. \cite{Domokos:2011dn} in several ways. Just like in ref. \cite{Cappiello:2010tu}, the authors do not include the effects of the dimension 5 bulk operator that communicates chiral symmetry breaking to the tensor sector, even though they correctly point out that its effect should be included. Secondly, in order to fix the degeneracies resulting from this truncation to a free field theory, they let the mass of the bulk $B_{\mu\nu}$ field take different values, so that the field is dual to operators of dimension $\Delta$ different than 3. While it is true, as we pointed out above, that using matching to free field theory is not a well justified procedure, it is at least a testable assumption and so far has met with surprisingly large phenomenological success. If one abandons this, one should not just treat the mass of the $B_{\mu \nu}$ field as a new free parameter, but also the five-dimensional gauge coupling, the mass and normalization of the bulk scalar field as well as the normalization of the $B_{\mu \nu}$ kinetic term, all of which affect the correlation functions in the boundary theory. In this case the model loses virtually all predictive power. Given the surprising accuracy with which the hard wall model so far has predicted particle masses, we believe it is premature to abandon the procedure of matching to UV correlators at this stage. We fix our parameters to reproduce the boundary expansion of a field dual to an operator of dimension $\Delta=3$.

In this paper we will explicitly carry out the calculation for the short distance behavior of the bulk correlation functions of $B_{\mu\nu}$. Comparing to the operator product expansion (OPE) of weakly coupled QCD we will indeed be able to completely fix all new coupling constants in the bulk. In fact, the set of conditions we obtain for the couplings is overdetermined and the fact that we can find values that allow us to reproduce all QCD correlation functions to leading order is a nice consistency check. The upshot is that this improved model has no new undetermined parameters. For now this serves as a proof of principle that this matching can be done. One has an improved hard wall model with no new free input parameters but several new predictions (masses and decay constants for the tensor and axial tensor mesons). Whether the phenomenological success of the model survives these additions will be a good test to what extent the underlying assumption of an effective description in five dimensions gives an accurate picture of real QCD. We  analyze the meson spectrum, we demonstrate that the cubic coupling indeed removes all the unwanted degeneracies of masses that were present in the case of a free $B_{\mu \nu}$, but unfortunately the meson spectrum that we observe does not match with what has been measured for QCD.

The organization of this paper is as follows: in section \ref{ss:opeqcd} we will review the short distance structure of the correlation functions involving the dimension 3 vector operators in QCD, as this is what we want to reproduce. In section \ref{ss:model} we present the improved holographic model. We derive equations of motion and the renormalized action in sections \ref{ss:equations} and \ref{ss:renormalization} respectively. In section \ref{ss:renormalization} we calculate the short distance correlation functions in the theory with massive quarks and extract the bulk coupling constants from comparing to QCD. As some of the correlation functions have the leading short distance terms proportional to the mass, in the chiral limit several correlators are dominated by the subleading term involving the chiral condensate. As our bulk Lagrangian at this stage is entirely determined, reproducing these correlators is a non-trivial check of our construction. We demonstrate that this indeed works out in section \ref{ss:matching}. In section \ref{ss:spectrum} we analyze the meson spectrum and the phenomenological issues that appear in the new model. In section \ref{ss:conclusions} we summarize our results.


\section{Correlation functions in QCD}\label{ss:opeqcd}


In two-flavor QCD, the relevant two-point functions in the vector sector are:
\begin{eqnarray}
&& \Pi^{\mu \nu,\, ab}_{VV}(q^2)=i\int d^4x\, e^{iqx}\bra \Omega T\{V^{\mu\, a}(x)V^{\nu\, b\, \dagger}(0)\}\ket \Omega,\\
&& \Pi^{\mu; \nu \rho,\, ab}_{VT}(q^2)=i\int d^4x\, e^{iqx}\bra \Omega T\{T^{\nu \rho\, a}(x)V^{\mu \dagger\, b}(0)\}\ket \Omega,\\
&& \Pi^{\mu \nu; \alpha \beta,\, ab}_{TT}(q^2)=i\int d^4x\, e^{iqx}\bra \Omega T\{T^{\mu \nu\, a}(x)T^{\alpha \beta \dagger\, b}(0)\}\ket \Omega,
\label{eq:tpoint}
\end{eqnarray}
where $V^{\mu\,a}(x)=\bar{\psi}(x)\gamma^{\mu}\tau^a\, \psi(x)$ and $T^{\mu \nu\, a}(x)=\bar{\psi}(x)\sigma^{\mu \nu}\tau^a\,\psi(x)$ are the vector and tensor isospin triplet currents respectively, and $\ket \Omega$ is the non-perturbative vacuum. We choose a normalization for the isospin generators such that $\tr(\tau^a \tau^b)=\frac{1}{2}\delta^{ab}$. The two-point functions above have the following kinematic structure:
\begin{eqnarray}
\Pi_{VV}^{\mu\nu,\, ab}(q^2) & = & \delta^{ab}(q^\mu q^\nu-q^2 \eta^{\mu\nu}) \Pi_{VV}(q^2), \\
\Pi_{TT}^{\mu\nu;\alpha\beta,\, ab}(q^2) & = & \delta^{ab}\Pi_{TT}^+(q^2) F_+^{\mu\nu;\alpha\beta} +\delta^{ab}\Pi_{TT}^-(q^2) F_-^{\mu\nu;\alpha\beta},\\
\Pi_{VT}^{\mu;\nu\rho,\, ab}(q^2) & = & i\delta^{ab}(\eta^{\mu\nu} q^\rho-\eta^{\mu \rho} q^\nu) \Pi_{VT}(q^2),
\label{eq:kin}
\end{eqnarray}
where, defining the projector $q^2P_{\mu\nu}=q^2 \eta_{\mu\nu}-q_\mu q_\nu$,
\begin{eqnarray}
 P_{[\mu}^{\alpha}P_{\nu]}^{\beta}=\frac{1}{q^2}F_{+\mu \nu}^{\alpha \beta};
\label{eq:pparity}
\end{eqnarray}
projects onto positive parity. Its counterpart is
\begin{equation}
 F^{\mu \nu;\alpha \beta}_-=F_+^{\mu \nu;\alpha \beta}-q^2(\eta^{\mu\alpha}\eta^{\nu\beta}-\eta^{\mu\beta}\eta^{\nu\alpha}),
\end{equation}
so $F_- \sim F_+-1$ is the negative parity projector. Notice that the sign of $F_-$ is chosen so it is actually minus the projector
\begin{equation}
F_-^{\mu\nu;\alpha\beta}=-(q^\nu q^\beta\eta^{\mu\alpha}+q^\mu q^\alpha \eta^{\nu\beta}-q^\nu q^\alpha\eta^{\mu\beta}-q^\mu q^\beta\eta^{\nu\alpha}).
\end{equation}
We also have:
\begin{equation}
(\delta_\mu^\alpha \delta_\nu^\beta-\delta_\mu^\beta \delta_\nu^\alpha)-P_{[\mu}^{\ \alpha} P_{\nu]}^{\ \beta}  = -\frac{1}{q^2} {F_-}_{\mu\nu}^{\ \ \alpha\beta} ,
\label{eq:nparity}
\end{equation}
with
\begin{equation}
{F_\pm}^{\mu\nu}_{\ \ \alpha\beta} {F_\pm}^{\alpha\beta}_{\ \ \sigma\rho} = \pm 2 q^2 {F_\pm}^{\mu\nu}_{\ \ \sigma\rho}.
\label{eq:ident}
\end{equation}
In the large-$N_c$ limit the two-point functions above are saturated by single-particle exchange of an infinite number of stable mesons, in this approximation to real QCD we can write, up to subtractions, the two-point functions above as:
\begin{eqnarray}
 &&\Pi_{VV}(q^2)=\sum_n\frac{f_{\rho,n}^2}{M_{\rho,n}^2-q^2};\,\,\,\,\,\,
\Pi_{TT}^-(q^2)=\sum_n\frac{(f_{\rho,n}^{T})^2}{M_{\rho,n}^2-q^2}\\
&&\Pi_{TT}^+(q^2)=\sum_n\frac{f_{b,n}^2}{M_{b,n}^2-q^2};\,\,\,\,\,\,\Pi_{VT}(q^2)=\sum_n\frac{f_{\rho,n}f_{\rho,n}^T}{M_{\rho,n}^2-q^2}\nonumber
\label{eq:spectral}
\end{eqnarray}
with the decay constants defined as:
\begin{eqnarray}
 &&\bra \Omega V_{\mu}^a\ket{\rho_n^b(p,\lambda)}=M_{\rho,n}\delta^{ab}f_{\rho,n}\epsilon_{\mu}(p,\lambda),\\
&&\bra \Omega T_{\mu \nu}^a\ket{\rho_n^b(p,\lambda)}=i\delta^{ab}f_{\rho,n}^T[p_{\mu}\epsilon_{\nu}(p,\lambda)-p_{\nu}\epsilon_{\mu}(p,\lambda)],\\
&&\bra \Omega T_{\mu \nu}^a\ket{b_n^b(p,\lambda)}=i\delta^{ab}f_{b,n}\varepsilon_{\mu \nu \alpha \beta}p^{\alpha}\epsilon^{\beta}(p,\lambda).
\end{eqnarray}
As it is made explicit by the notation above, the current $V^{\mu}$ produces vector mesons ($J^{PC}=1^{--}$) like the $\rho$, while the tensor operator $T^{\mu \nu}$ produces both vector mesons and their even-parity partners ($J^{PC}=1^{+-}$), like the $b_1$ meson.
For large Euclidean momentum $Q^2=-q^2 \rightarrow \infty$ contributions to these correlators can be organized according to the operator product expansion (OPE), with a leading perturbative contribution plus an expansion on the several vacuum condensates, $\langle \bar{\psi}\psi\rangle$, $\langle \alpha_s G^2\rangle$, etc., that capture the non-perturbative effects. This was originally done for three colors in refs. \cite{Shifman:1978bx,Shifman1979519,Govaerts1987706}. Expressions for a general number of colors can also be found in refs. \cite{Reinders:1984sr,PhysRevD.26.2430,Cata:2008zc}. To leading order we have:
\begin{eqnarray}
\label{eq:vvope}
 &&\lim_{Q^2 \to \infty}\Pi_{VV}(Q^2)=-\frac{N_c}{24\pi^2}\log\frac{Q^2}{\mu^2}+\mathcal{O}\left(\frac{\alpha_s}{Q^4}\right),\\
\label{eq:ttope}
 &&\lim_{Q^2 \to \infty}\Pi_{TT}^\pm(Q^2)=-\frac{N_c}{48\pi^2}\log\frac{Q^2}{\mu^2}\mp\frac{N_c}{8\pi^2}\frac{ m^2}{Q^2}\log\frac{Q^2}{\mu^2}+\mathcal{O}\left(\frac{\alpha_s}{Q^4}\right),\\
\label{eq:vtope}
&&\lim_{Q^2 \to \infty}\Pi_{VT}(Q^2)=\frac{N_c}{16\pi^2}m\log\frac{Q^2}{\mu^2}-\frac{\langle\bar{\psi}\psi\rangle}{Q^2}+\mathcal{O}\left(\frac{\alpha_s}{Q^4}\right),
\end{eqnarray}
where $m$ is the quark mass. This is the large momentum behaviour of the correlators that we will use to fix the free parameters of the five-dimensional action.


\section{Improved model of holographic QCD}\label{ss:model}


The model we consider is an extension of the hard wall model of ref. \cite{Erlich:2005qh}, but it can be generalized to other holographic QCD models like the soft wall model of ref. \cite{Karch:2006pv}. We use a five-dimensional geometry to describe the dynamics of four-dimensional QCD with a large number of colors $N_c\to\infty$. The metric is that of $AdS_5$ with a radius $\ell$, we choose a mostly minus signature and work with the coordinate system
\begin{equation}
ds^2=g_{MN}dx^M dx^N=\frac{\ell^2}{z^2}\left(-dz^2+\eta_{\mu\nu} dx^\mu dx^\nu\right)\,.
\end{equation}
In these coordinates the boundary is at $z=0$. In order to recover some of the physics of confinement, we introduce a cutoff in the radial coordinate $z_m$. Since the radial coordinate maps to a renormalization group scale in the dual theory, with $z=0$ corresponding to the UV, the cutoff $1/z_m$ can be interpreted as an IR scale where the theory becomes confining.

We introduce a set of fields $\phi(x,z)$ in the five-dimensional theory that are dual to mesonic operators ${\cal O}(x)$ in the field theory with conformal dimensions $\Delta \leq 3$ and spin $J\leq 1$. In previous works this was done considering scalar and vector fields. This included both scalar and pseudoscalar mesons, as well as vector mesons $1^{--}$ and axial vector mesons $1^{++}$, but the full set of vector mesons include also $1^{+-}$ states, that were missing in the original formulation. These can be included by considering a complex two-index antisymmetric field $B_{\mu\nu}$, or two-form for short.

The five-dimensional theory has a $U(2)_L\times U(2)_R$ gauge symmetry, that maps to the global flavor symmetry of two-flavor QCD. The fields $A_{L\,\mu}$ and $A_{R\,\mu}$ will be the associated gauge bosons, while the complex fields $X$ and $B_{\mu\nu}$ are in a bifundamental representation
\begin{equation}
X\longrightarrow U_L X U_R^\dagger, \ \ \ B_{\mu\nu}\longrightarrow U_L B_{\mu\nu} U_R^\dagger.
\end{equation}
The map between operators and fields can be summarized as:
$$
\begin{array}{llcc}
4D: {\cal O}(x) & 5D: \phi(x,z) &  \Delta & m_\phi^2\ell^2 \\ \hline
\sqrt{2}\overline{\psi}_L  \gamma_\mu \tau^a \psi_L & A^a_{L\,\mu} &  3 & 0 \\
\sqrt{2}\overline{\psi}_R  \gamma_\mu \tau^a \psi_R & A^a_{R\,\mu} &  3 & 0 \\
\overline{\psi}_L^\alpha \psi_R^\beta & X^{\alpha\beta} &  3 & -3 \\
\overline{\psi}_L^\alpha  \sigma_{\mu\nu} \psi_R^\beta & B^{\alpha\beta}_{\mu\nu} &  3 & 1 \\
\end{array}
$$
Where $m^2_\phi\ell^2$ is the mass of the field. We have chosen masses such that the conformal dimension $\Delta$ of the dual operator matches with its free value. Although quantum corrections will change the conformal dimension of operators in the IR, QCD is a free theory in the UV and is in this regime where we will do the matching with our model, hence our choice of masses both for the scalar and the two-form field. We can also form the real combinations
$$
\begin{array}{llcc}
4D: {\cal O}(x) & 5D: \phi(x,z) &  \Delta & m_\phi^2\ell^2 \\ \hline
\overline{\psi}  \gamma_\mu \tau^a \psi & V_{\mu}^a=(A^a_{R\,\mu}+ A^a_{L\,\mu})/\sqrt{2} &  3 & 0 \\
\overline{\psi}  \gamma_\mu\gamma_5 \tau^a \psi & A_{\mu}^a= (A^a_{R\,\mu}- A^a_{L\,\mu})/\sqrt{2} &  3 & 0 \\
\overline{\psi}^\alpha   \psi^\beta & X_+^{\alpha\beta}=X^{\alpha\beta}+  X^{\dagger\,\alpha\beta}&  3 & -3 \\
i\,\overline{v}^\alpha \gamma_5  \psi^\beta & X_-^{\alpha\beta}=i\left(X^{\alpha\beta}-  X^{\dagger\,\alpha\beta}\right)&  3 & -3 \\
\frac{1}{\sqrt{2}}\overline{\psi}^\alpha  \sigma_{\mu\nu} \psi^\beta & B_{+\, \mu\nu}^{\alpha\beta}=(B^{\alpha\beta}_{\mu\nu}+B^{\dagger\,\alpha\beta}_{\mu\nu})/\sqrt{2} &  3 & 1 \\
\frac{i}{\sqrt{2}}\,\overline{\psi}^\alpha  \sigma_{\mu\nu} \gamma_5 \psi^\beta & B_{-\, \mu\nu}^{\alpha\beta}=i\left(B^{\alpha\beta}_{\mu\nu}-B^{\dagger\,\alpha\beta}_{\mu\nu} \right)/\sqrt{2} &  3 & 1 \\
\end{array}
$$
Although the flavor representation is correct, in four dimensions a complex two-form has too many degrees of freedom, the reason is that the tensor operators are not all independent, but satisfy the duality condition \eqref{eq:dualitycond}. This relation implies that the complex two-form has to be imaginary anti self-dual.

Defining $F_L$ and $F_R$ as the field strengths of $A_L$ and $A_R$, $H=dB-iA_L\wedge  B+iB\wedge  A_R$ as the three-form field strength of $B$, and $DX= dX-i A_L X+i X A_R$ as the covariant derivative of $X$, the action takes the form
\begin{align}
\notag S&=\int d^5 x\sqrt{-g}\,{\rm tr}\,\Big[-\frac{1}{4g^2_5}(F^2_L+F^2_R)+g_X^2\big(\left|D X\right|^2+3 \left|X\right|^2 \big) -2g_B\left(i\frac{1}{6}(B\wedge H^\dagger-B^\dagger \wedge H)+m_B |B|^2\right)+\\ &+\frac{\lambda}{2}(X^{\dagger}F_LB+BF_RX^{\dagger}+h.c.)\Big]\,.
\label{eq:action1}
\end{align}
The trace is taken over the gauge indices. The factors of the $AdS$ radius $\ell$ have been absorbed in the coupling constants or the masses.

Let us comment on the different terms. The first term is the kinetic action of the gauge fields, its coefficient was fixed in the original hard wall model comparing the expansion of the holographic vector-vector correlation function at large Euclidean momentum with the OPE of QCD \cite{Erlich:2005qh}. The result was
\begin{equation}\label{eq:g5}
\frac{1}{g_5^2} = \frac{N_c}{12\pi^2}\,.
\end{equation}
The second term is the scalar action, that is usually canonically normalized, which can be achieved by rescaling $g_X X \rightarrow X$ in \eqref{eq:action1}.
The asymptotic value of the scalar field close to the boundary $z=0$ determines the quark mass $m$ and the condensate $\left\langle \overline{\psi} \psi\right\rangle$ in the dual theory. With canonical normalization, $g_X$ now appears in this relation:
\begin{equation}\label{eq:scalar}
X=\frac{1}{2} \left(g_X m \, z+ \frac{\left\langle \overline{\psi} \psi\right\rangle}{g_X} z^3\right) \mathbf{1}_{2\times 2}\equiv \frac{g_X}{2} v(z) \mathbf{1}_{2\times 2}\,.
\end{equation}
The reason that there are not two independent normalization constants in this relation is that the expectation value of the mass operator $\left\langle \overline{\psi} \psi\right\rangle$ can be obtained from varying the on-shell action with respect to $m$.
The value of $g_X$ was determined in ref. \cite{Cherman:2008eh} comparing the holographic scalar-scalar correlation function at large Euclidean momentum with the OPE of QCD,
\begin{equation}\label{eq:gX}
g_X^2=\frac{N_c}{4\pi^2}\,.
\end{equation}
The third term is the action of the two-form field. The kinetic term in the action has been replaced by a Chern-Simons term, and the mass $m_B$ is a free parameter. We will see latter that the right self-duality condition for the two-form field can be derived from the equations of motion of this action by fixing the mass. Then, following the usual procedure, we will compute the holographic tensor-tensor correlator, expand it at large Euclidean momentum, and match with the OPE of QCD. The last term is the most general gauge-invariant term of dimension five or less that couple isospin triplet vector mesons and preserve parity and charge conjugation in the dual theory \cite{Domokos:2011dn}.

We can rewite the interaction term in \eqref{eq:action1} using real fields
\begin{equation}\label{eq:veccoupling}
{\rm tr}\,\left(X^{\dagger}F_LB+BF_RX^{\dagger}+h.c.\right)=\frac{1}{2}{\rm tr}\,\Big( X_+ \big(\{ F_V,B_+\}+i [F_A,B_-] \big)+ X_- \big(\{ F_V,B_-\}-i [F_A,B_+] \big)\Big)\,.
\end{equation}
The term \eqref{eq:veccoupling} determines how chiral symmetry breaking affects to the isospin triplet $1^{--}$ vector mesons. Without this coupling the spectrum will be determined by the equations of motion of the $V$ field, but once we introduce it, the $B_+$ field and the $V$ field are coupled.

Using the expression \eqref{eq:scalar} for the background scalar field and taking the trace in the action \eqref{eq:action1}, we get in the vector sector
\begin{align}
\notag S_V&=\int d^5 x \sqrt{-g}\Big[-\frac{1}{4g^2_5}\sum_{i=V,A} F_{i\,MN}F^{i\,MN}+\frac{g_B}{3}\varepsilon^{MNLPQ}\big(B_{-MN}H_{+LPQ}-B_{+MN}H_{-LPQ}\big)-\\ &-g_B m_B\sum_{\alpha=+,-}B_{\alpha MN}B_{\alpha}^{MN}+\frac{\lambda}{2} v(z)F_{V\,MN}B_+^{MN}\Big]\,,
\label{eq:action2}
\end{align}
where we have suppressed gauge indices. In total we have introduced three new parameters, $g_B$, $m_B$ and $\lambda$. We will now fix $m_B$ imposing the self-duality condition and $g_B$ and $\lambda$ using the matching with the OPE of QCD. The only free parameters left in the model are the mass $m$, the condensate $\sigma=\left\langle \overline{\psi} \psi\right\rangle/g_X^2$ and the IR scale $1/z_m$. Although by introducing the $B$ field we have added a new sector of vector mesons and therefore of masses and decay constants we can compare the model with, we have not increased the number of free parameters. Notice that the axial sector, involving the fields $X$  and $A$ is untouched.


\section{Equations of motion}\label{ss:equations}


Our next step is to calculate the equations of motion for the fields $B_\pm^{\mu z}$,$B_\pm^{\mu \nu}$, and $V^{\mu}$ from the action \eqref{eq:action2}. We will write explicitly all the factors involving the radial coordinate and raise and lower indices with the flat metric, using $g_{MN}=\frac{1}{z^2} \eta_{MN}$. Greek letters for the indices will refer to the flat Minkowsky directions and capitalized italic letters will include the radial direction $z$.
Let us consider first the case with no interaction, $\lambda=0$. From \eqref{eq:action2} we get the equations of motion for the components of the two-form
\begin{eqnarray}
 \pm \frac{1}{3}\varepsilon^{MNLPQ}H_{\mp MNL}+m_B B^{PQ}_{\pm}=0.
\label{eq:eom1}
\end{eqnarray}
Including explicit powers of z we have:
\begin{eqnarray}
\label{eq:eom2a}
&& \pm \epsilon^{z\alpha \beta \mu\nu}H_{\mp z\alpha \beta}+\frac{m_B}{z}B^{\mu \nu}_{\pm}=0,\\
\label{eq:eom2b}
&& \pm \epsilon^{z\alpha \beta \gamma \mu}H_{\mp \alpha \beta \gamma}+\frac{3m_B}{z}B^{\mu z}_{\pm}=0,
\end{eqnarray}
with the definition $\epsilon_{z \alpha \beta \mu\nu}\equiv\epsilon_{\alpha \beta \mu \nu}$\,\,$\Longrightarrow$\, $\epsilon^{z \alpha \beta \mu\nu}=-\epsilon^{\alpha \beta \mu\nu}$\,\,. After some algebra, one can write the equations of motion as
\begin{eqnarray}
\label{eq:bmn}
&&  z\partial_z\big(zH^{z \mu\nu}_\pm\big)+z^2\partial_{\alpha}H_\pm^{\alpha \mu \nu}+\frac{m_B^2}{4}B_\pm^{\mu \nu}=0,\\
\label{eq:bz}
&&  \partial_{\alpha}H_\pm^{\alpha \mu z}+\frac{m_B^2}{4z^2}B_\pm^{\mu z}=0.
\end{eqnarray}

Note that equation \eqref{eq:eom1} is of the type $dB-*B=0$ which implies the constraint $d*B=0$. In components,
\begin{eqnarray}
\label{eq:c1}
&& \partial_{\mu}B_\pm^{\mu z}=0,\\
\label{eq:c2}
&& \partial_{\mu}B_\pm^{\mu \nu}+z\partial_z\frac{1}{z}B_\pm^{z \nu}=0.
\end{eqnarray}

Using \eqref{eq:c2} in \eqref{eq:bmn} and \eqref{eq:bz} we can eliminate $\partial_zB_\pm^{\nu z}$ from the former and $B_\pm^{\mu \nu}$ from the latter. Expanding solutions in Fourier modes of four-momentum  $q^\mu$ we get:
\begin{eqnarray}
&& z^2\partial_z^2B_\pm^{\mu \nu}+z\partial_zB_\pm^{\mu \nu}+\left(z^2q^2-\frac{m_B^2}{4}\right)B_\pm^{\mu \nu}=2izq^{[\mu}B_\pm^{\nu] z},\\
&& \partial_z^2B_\pm^{\mu z}-\frac{1}{z}\partial_zB_\pm^{\mu z}+\left(q^2+\frac{4-m_B^2}{4z^2}\right)B_\pm^{\mu z}=0.
\end{eqnarray}
Let us use the following decomposition
\begin{equation}\label{eq:bexp}
B_\pm^{\mu\nu} = i q^{[\mu}T^{\nu]}_\pm+i\epsilon^{\mu\nu\sigma\rho}q_\sigma \overline{T}_{\pm\, \rho}.
\end{equation}
If the two-form field is dual to an operator of conformal dimensions $\Delta=3$, then its expansion at small $z$ should be
\begin{align} \nonumber
& T_\pm^\mu=\frac{1}{z} \Tz_\pm^\mu+z \log z\To_\pm^\mu+z\Tt_\pm^\mu+\cdots \\
& {\overline{T}}_\pm^\mu=\frac{1}{z} {\tTz}_\pm^\mu+z \log z{\tTo}_\pm^\mu+z{\tTt}_\pm^\mu+\cdots \label{eq:texp}
\end{align}
This is possible if we set $m_B^2=4$. Setting $m_B=2$ and  using the original equations \eqref{eq:eom2a} and \eqref{eq:eom2b}, from the leading $\sim 1/z^2$ term we get the conditions
\begin{equation}\label{eqcond0}
{\tTz}_\mp^\mu=\pm \Tz_\pm^\mu\,.
\end{equation}
The conditions \eqref{eqcond0} above imply that
\begin{equation}\label{eq:boundB}
\bz_{+\, \mu\nu}+\frac{1}{2}\epsilon_{\mu\nu\alpha\beta}\bz_-^{\alpha\beta}=0.
\end{equation}
Let us define $b_{\mu\nu}=\bz_{+\,\mu\nu}-i\bz_{-\,\mu\nu}$. In terms of $b$, the condition above means it is imaginary anti-self-dual
\begin{equation}
b+i*b=0.
\end{equation}
The conjugate $b^\dagger$ is imaginary self-dual.

Let us now study the effect of adding to the action a kinetic term for the two-form field, with a relative coefficient $C>0$.
\begin{align}\label{eq:kinetic}
&& \Delta S=g_B C \int d^5 x \sqrt{-g}\Big[\sum_{\alpha=+,-}H_{\alpha MNL}H_{\alpha}^{MNL}\Big]\,,
\end{align}
Given arbitrary mass $m_B$, we find the following equations
\begin{equation}\label{eq:geneq}
C \nabla_L H_\pm^{\ LMN}\pm \frac{1}{3} \varepsilon^{MNABC}H_{\mp\,ABC}+m_B B_\pm^{\ MN}=0\,.
\end{equation}
Where $\nabla_L\equiv \frac{1}{\sqrt{-g}}\partial_L \sqrt{-g}$. Assuming that the leading term in the small-$z$ expansion of the two-form field is $B_\pm^{\ \mu\nu} \simeq z^\nu\bz_\pm^{\mu\nu}$, one finds that $\bz_\pm^{\mu\nu}$ satisfies an imaginary (anti) self-duality condition if
\begin{equation}
C \nu^2 \mp 2\nu-m_B=0.
\end{equation}
The upper sign corresponds to the condition \eqref{eq:boundB}. The derivation is valid if $m_B-C\nu^2\neq 0$, otherwise the leading term is a logarithm and the analysis is different. The equations for plus and minus components can be decoupled, giving a fourth order equation
\begin{equation}\label{eq:quartic}
(4+2 m_B C)\nabla_K H_\pm^{\ KPQ}+m_B^2 B_\pm^{\ PQ}+\frac{C^2}{2}\nabla_L\left[\left(g^{KP}g^{CQ}g^{DL}+(PQL)\right) \partial_K\left(g_{CU} g_{DV} \nabla_S H_\pm^{\ SUV} \right)  \right]=0.
\end{equation}
Where by $(PQL)$ we denote permutations with a minus sign if they are odd with respect to the first term. To leading order in $z$, the equation for the spacetime components $B_\pm^{\ \mu\nu}$ imposes a constraint on $\nu$ in the form of a quartic equation
\begin{equation}
C^2 \nu^4-(4+2m_BC)\nu+m_B^2=0.
\end{equation}
We can rewrite this equation as
\begin{equation}
(C\nu^2+2\nu-m_B)(C\nu^2-2\nu-m_B)=0.
\end{equation}
Therefore, when the solution is (anti) self-dual the quartic equation is automatically satisfied. This shows that we can always impose the right self-duality condition, although we can also have solutions with opposite self-duality conditions if we change $\nu$, we have to set those to zero by hand. Instead, we will drop the kinetic term, we will show that it does not affect to properties of the model like the meson spectrum, although it can affect to correlation functions because it contributes to the boundary action.

If we set $C=0$, then the equations \eqref{eq:quartic} become
\begin{equation}\label{eq:quartic2}
\nabla_K H_\pm^{\ KPQ}+\frac{m_B^2}{4} B_\pm^{\ PQ}=0.
\end{equation}
Let us now do the following trick, we can rewrite \eqref{eq:geneq} as
\begin{equation}\label{eqaux}
C\left[ \nabla_L H_\pm^{\ LMN}+\frac{\tilde{m}_B^2}{4}B_\pm^{\ MN}\right]\pm \frac{1}{3} \varepsilon^{MNABC}H_{\mp\,ABC}+\left(m_B-\frac{\tilde{m}_B^2 C}{4}\right) B_\pm^{\ MN}=0\,.
\end{equation}
Then, if
\begin{equation}\label{eq:massshift}
m_B=\tilde{m}_B+\frac{\tilde{m}_B^2 C}{4},
\end{equation}
the solutions to the $C=0$ decoupled equation \eqref{eq:quartic2} with mass $\tilde{m}_B$ would make the term that multiplies $C$ in the first bracket vanish, so one would recover the $C=0$ equations again. Now let us fix the asymptotic expansion \eqref{eq:texp} ($\nu=-1$) and impose the self-duality condition \eqref{eq:boundB}. By fixing the mass to the value $m_B=2+C$ we can solve the system with $C\neq 0$ using the solutions for $C=0$.\footnote{We can consider $C<0$ by changing the overall sign of the action.} If one examines the equations involving the interaction term below one sees that the same is true if the coupling is rescaled appropriately. An exception to this rule may be the case $m_B=0$, where there is an additional gauge invariance associated to the two-form field $\delta B_{MN}=\partial_{[M} \Lambda_{N]}$ and the separation in two parts of the equations of motion involves introducing gauge non-invariant terms. It also coincides with the case $\nu=0$, where the leading solution is logarithmic. We will neglect this case and set $C=0$ from now on.

We now consider the interaction term, it does not affect to the leading asymptotic behavior, so the value of $m_B=2$ is not changed. The equations of motion of the $B_+$ two-form are
\begin{eqnarray}
\label{eq:fb}
&&z\partial_{\alpha}H_{+}^{\alpha \mu \nu}+\partial_zzH_{+}^{z \mu \nu}+\frac{B_{+}^{\mu \nu}}{z}=\frac{\lambda}{8g_B}v(z)\frac{F^{\mu \nu}_{V}}{z},\\
\label{eq:fz}
&& \partial_{\alpha}zH^{\alpha \nu z}_{+}+\frac{B^{\nu z}_{+}}{z}=\frac{\lambda}{8 g_B}\frac{v(z)}{z}F^{\nu z}_{V},\\
&&z\partial_{\alpha}H_{-}^{\alpha \mu \nu}+\partial_zzH_{-}^{z \mu \nu}+\frac{B_{-}^{\mu \nu}}{z}=-\frac{\lambda}{8g_B}v'(z)F_{V \alpha \beta}\epsilon^{\alpha \beta \mu \nu},\\
&& \partial_{\alpha}zH^{\alpha \nu z}_{-}+\frac{B^{\nu z}_{-}}{z}=0.
\end{eqnarray}
The equations of motion of the vector fields and the constraints for the two-form fields are
\begin{eqnarray}
\label{eqv1} &&\partial_z\frac{1}{z}B_{+}^{\nu z}-\frac{1}{z}\partial_{\mu}B_{+}^{\mu \nu}=\frac{\lambda}{8 g_B}\left[\partial_z\frac{v(z)}{z}F_{V}^{\nu z}-\frac{v(z)}{z}\partial_{\mu}F_{V}^{\mu \nu}\right],\\
\label{eqv2} && \partial_z\frac{1}{z}F_{V}^{\nu z}-\frac{1}{z}\partial_{\mu}F_{V}^{\mu \nu}=\lambda g_5^2\left[\partial_z\frac{v(z)}{z}B_{+}^{ \nu z}-\frac{v(z)}{z}\partial_{\mu}B_{+}^{\mu \nu}\right],\\
\label{eqv3} &&\partial_z\frac{1}{z}B_{-}^{\nu z}-\frac{1}{z}\partial_{\mu}B_{-}^{\mu \nu}=0,\\
\label{eqv4} && \partial_z\frac{1}{z}F_{A}^{\nu z}-\frac{1}{z}\partial_{\mu}F_{A}^{\mu \nu}=g_5^2g_X^2\frac{v^2(z)}{z^3}A^{\nu}.
\end{eqnarray}
We now expand in Fourier modes and simplify the system to:
\begin{eqnarray}
\label{eq:vecb}
 &&\partial_zB^{\nu z}_+-\frac{f(z)}{g(z)}B^{\nu z}_+-iq_{\mu}B^{\mu \nu}_+=\frac{\lambda}{8g_B}\frac{v'(z)}{g(z)}\partial_zV^{\nu}\\
\label{eq:vecv}
&&\partial_z^2V^{\nu}-\frac{f(z)}{g(z)}\partial_zV^{\nu}+q^2V^{\nu}=\frac{\lambda g_5^2v'(z)}{g(z)}B_+^{\nu z}\\
\label{eq:em1}
 &&\partial_zB^{\nu z}_--\frac{1}{z}B^{\nu z}_--iq_{\mu}B^{\mu \nu}_-=0\\
\label{eq:em2}
&&\partial_z^2A^{\nu}-\frac{1}{z}\partial_zA^{\nu}+q^2A^{\nu}=g_5^2g_X^2\frac{v^2(z)}{z^2 }A^{\nu}
\end{eqnarray}
Where $f(z)=\frac{1}{z}+\chi\,zv(z)(\frac{v(z)}{z})'$, $g(z)=1-\chi\,v(z)^2$ and $\chi=\lambda^2g_5^2/(8 g_B)$.
Note the structure of the equations, we can understand how the fields mix among themselves by considering parity conservation:
the vector mode $V^\mu$ (negative parity) couples to its vector partner $B^{\nu z}_+$ \eqref{eq:vecv} and to the tensor component of negative parity \eqref{eq:fb}. We can further decouple the system of equations above:
\begin{eqnarray}
\label{eq:bzp}
&& \left[\partial_z^2-\frac{f(z)}{g(z)}\partial_z-(C_1(z)-q^2)\right]B_+^{\nu z}=-\frac{\lambda}{8 g_B}\left[C_2(z)V'^{\nu}+\frac{v'(z)q^2}{g(z)}V^{\nu}\right],\\
\label{eq:bzm}
&& \left[\partial_z^2-\frac{1}{z}\partial_z+q^2\right]B_-^{\nu z}=0,
\end{eqnarray}
where
\begin{eqnarray}
&& C_1(z)=\partial_z\frac{f(z)}{g(z)}+\frac{1}{z^2}+\frac{\chi\, v'^2}{g(z)^2}, \\
&& C_2(z)=\frac{v(z)}{z^2}-\partial_z\frac{v'(z)}{g(z)}-v'(z)\frac{f(z)}{g^2(z)}.
\end{eqnarray}
The equations for the tensor components of the two-form field are
\begin{eqnarray}
&&\left[ z\partial_zz\partial_z-1+z^2q^2\right]B_+^{\mu \nu }=-i\frac{\lambda}{8 g_B}\left[v(z)q^{[\mu}V^{\nu]}-\frac{z^2v'(z)}{g(z)}\partial_zq^{[\mu}V^{\nu]}\right]+
\left(z^2\frac{f(z)}{g(z)}+z\right)iq^{[\mu}B_+^{\nu] z},
\label{eq:tensorp}\\
&&\left[ z\partial_zz\partial_z-1+z^2q^2\right]B_-^{\mu \nu}=-i\frac{\lambda}{8 g_B} zv'(z)q_{[\alpha}V_{\beta]}\epsilon^{\alpha \beta \mu \nu}+
2z iq^{[\mu}B_-^{\nu] z}.
\label{eq:tensorm}
\end{eqnarray}
Our final equations of motion can then be divided in the vector \eqref{eq:vecv},  axial vector \eqref{eq:em2},  and two-form components  \eqref{eq:bzp}, \eqref{eq:bzm}, \eqref{eq:tensorp} and \eqref{eq:tensorm}. A more convenient grouping is in four decoupled sets: $\{A^\mu\}$, $\{V^\mu, B_+^{\mu z}, (\delta^\mu_\alpha-{ P}^\mu_\alpha) B_+^{\alpha \nu},{ P}^\mu_\alpha { P}^\nu_\beta B_-^{\alpha\beta} \}$, $\{{ P}^\mu_\alpha { P}^\nu_\beta B_+^{\alpha\beta}\}$ and $\{ B_-^{\mu z}, (\delta^\mu_\alpha-{P}^\mu_\alpha) B_-^{\alpha \nu}\}$. Normalizable solutions of the first two sets correspond to $1^{++}$ and $1^{--}$ mesons in the dual theory, respectively. The last two are not independent since they are coupled in the original system of first order equations \eqref{eq:eom2a}, \eqref{eq:eom2b}, and normalizable solutions correspond to $1^{+-}$ mesons. Notice that  $b_+^{\mu\nu}\equiv z{P}^\mu_\alpha {\cal P}^\nu_\beta B_+^{\alpha\beta}$ satisfies the same equation as $B_-^{\mu z}$, and that this one is the same as the equation for vector mesons \eqref{eq:vecv} in the absence of the interaction term $\lambda=0$. Therefore, the interaction lifts the degeneracy between $1^{--}$ and $1^{+-}$ mesons.

\subsection{Boundary expansion}

We now proceed to do a Frobenius expansion of solutions close to the boundary at $z=0$. This will be useful for both the calculation of renormalized two-point functions and the calculation of the meson spectrum. Using \eqref{eq:bexp} in \eqref{eq:eom2a}, we find the conditions
\begin{align}\nonumber
&\mp  (\partial_z T_{\mp\,\mu}+B_{\mp\,\mu z})+\frac{1}{z}\overline{T}_{\pm\,\mu}=0,\\ \nonumber
&  \partial_z\overline{T}^\mu_{-}+\frac{1}{z} T^\mu_{+}=\tlp\frac{v}{z}V^\mu,\\
& - \partial_z\overline{T}^\mu_{+}+\frac{1}{z} T^\mu_{-}=0. \label{eq:boundcond1}
\end{align}
And from \eqref{eq:eom2b} we have
\begin{align} \notag
&-q^2 \overline{T}^\mu_-+\frac{1}{z} B^{\mu z}_+=\tlp \frac{v}{z} \partial_z V^\mu,\\
& q^2 \overline{T}^\mu_++\frac{1}{z} B^{\mu z}_-=0.\label{eq:boundcond2}
\end{align}

The expansion of the vector components at small $z$ is given by \eqref{eq:texp} and
\begin{align} \nonumber
& A_\pm^\mu = \az_\pm^\mu+z^2\log z\ao_\pm^\mu+z^2\at_\pm^\mu+\cdots, \\
& B^{\mu z}_\pm= \bz_\pm^\mu+z^2 \log z\bo_\pm^\mu+z^2\bt_\pm^\mu+\cdots, \label{eq:vecexp}
\end{align}
where we have defined $A_+^\mu=V^\mu$ and $A_-^\mu=A^\mu$. Expanding \eqref{eq:boundcond1}, \eqref{eq:boundcond2}, \eqref{eqv2} and \eqref{eqv4} for small $z$ we find a set of conditions that allows us to solve for the coefficients of the logarithmic terms and give us a relation between the leading terms, dual to sources in the field theory. In particular we recover the imaginary self-duality condition \eqref{eqcond0} for the components of the two-form field. Defining $\tilde{\lambda}_+=\lambda/(8 g_B)$,  $\tilde{\lambda}_-=0$, $q_+^2=q^2$ and $q_-^2=q^2-g_X^2 g_5^2m^2$, we can write them in a compact form
\begin{align}
\notag &\bz_\pm^\mu =\pm q^2\Tz_\pm^\mu, &{\tTz}_\mp^\mu=\pm \Tz_\pm^\mu, \\
\notag &\To^\mu_\pm =\frac{q^2}{2}\Tz_\pm^\mu-\frac{\tlpm}{2}m\az_\pm^\mu, &{\tTo}_\mp= \mp\frac{q^2}{2}\Tz_\pm^\mu\pm\frac{\tlpm}{2}m\az_\pm^\mu, \\
&{\tTt}_\mp^\mu=\mp\Tt_\pm^\mu\pm\frac{1}{2}(q^2\Tz_\pm^\mu+\tlpm m \az_\pm^\mu) &\ao_\pm^\mu=-\frac{1}{2}(q_\pm^2\az_\pm^\mu-g_5^2 \lambda_\pm m q^2\Tz_\pm^\mu).\label{eqcond}
\end{align}


\section{Holographic renormalization}\label{ss:renormalization}


We will follow the usual holographic procedure to compute correlation functions, deriving the on-shell action with respect to the sources of dual operators. The action usually diverges, so we will introduce a cutoff at a small value of the radial coordinate $z=\varepsilon$ to regularize it. We will introduce counterterms following the usual prescription \cite{deHaro:2000xn} to make the action finite before removing the cutoff $\varepsilon\to 0$.

The on-shell regularized action is
\begin{equation}
S_{o.s.}=\int d^4 x\sqrt{-g} \left[\frac{1}{2 g_5^2}  g^{zz} g^{\mu\nu} \sum_{a=\pm}F_{\pm\,z\mu} A_{\pm\,\nu}-\lambda v g^{zz} g^{\mu\nu} B_{+,z\mu} A_{+,\nu} \right]_{z=\varepsilon}+S_{CS}.
\end{equation}
Where we have introduced the cutoff $\varepsilon$ and the overall sign corresponds to taking the lower limit in the $z$ integral. $S_{CS}$ is the contribution of the two-form Chern-Simons action. The action has the form
\begin{equation}
S_{CS}= -g_B \int d^5 x\,\frac{i}{3}\tr\left(B\wedge H^\dagger-B^\dagger \wedge H\right)=\frac{g_B}{3}\int d^5 x \varepsilon^{MNLPQ}\big(B_{-MN}H_{+LPQ}-B_{+MN}H_{-LPQ}\big).
\end{equation}
A variation gives, to leading order,
\begin{align}
&\delta S_{CS} = -\frac{g_B}{3}\int d^4 x\, \varepsilon^{\mu\nu\alpha\beta}\big(B_{-\mu\nu}\delta B_{+\alpha\beta}-B_{+\mu\nu}\delta B_{-\alpha\beta}\big).
\end{align}
Where the fields are evaluated at the cutoff $z=\varepsilon$. The condition \eqref{eq:boundB} implies that we cannot vary $B_+$ and $B_-$ independently, but since $B_++* B_-=0$ at the boundary, we should treat $B_+-*B_-$ as the variable boundary value. In order to have a consistent variational principle we need to add a boundary term to the action, of the form
\begin{equation}
S_0=\frac{2g_B}{3}\int d^4 x \,\sqrt{-\gamma}\tr\left( \gamma^{\mu\alpha}\gamma^{\nu\beta}B^\dagger_{\mu\nu} B_{\alpha\beta}\right),
\end{equation}
where the indices are raised with the induced boundary metric $\gamma_{\mu\nu}=\varepsilon^{-2} \eta_{\mu\nu}$. The variation of this term, with expicit $\varepsilon$ factors, is
\begin{equation}
\delta S_0=\frac{2g_B}{3}\int d^4 x \big(B_+^{\ \mu\nu}\delta B_{+\mu\nu}+B_{-}^{\ \mu\nu}\delta B_{-\mu\nu}\big)\,.
\end{equation}
The sum of the two variations gives
\begin{equation}
\delta (S_{CS}+S_0)=\frac{2g_B}{3}\int d^4 x \Big(B_+^{\ \mu\nu}+\frac{1}{2}\epsilon^{\mu\nu\alpha\beta} B_{-\,\alpha\beta}\Big)\delta\Big( B_{+\mu\nu}-\frac{1}{2}\epsilon_{\mu\nu}^{\ \ \sigma\rho} B_{-\,\sigma\rho}\Big)\,.
\end{equation}
This gives a consistent variational principle.

The total on-shell regularized action is
\begin{equation}
S_{o.s.}=\int d^4 x \ \left[-\frac{1}{2 g_5^2}\, \frac{1}{z} \sum_{a=\pm} \partial_z A_a^\mu A_{a\,\mu}+\lambda \frac{v}{z} B_{+\,z}^{\ \mu} V_{\mu}+\frac{g_B}{3}\sum_{a=\pm} B_{a\,\mu\nu}B_a^{\mu\nu} \right]_{z=\varepsilon}.
\end{equation}
The bulk contribution vanishes on-shell.

Expanding for small $\varepsilon$ we find that, for Fourier modes of momentum $q_\mu$,
\begin{eqnarray}
\notag  \frac{1}{z} \partial_z A_\pm^\mu A_{\pm\,\mu} & \sim & (2 \at_{\pm\,\mu}+\ao_{\pm\,\mu}) \az_\pm^\mu +\\
& & +2 \log(Q \varepsilon) \ao_{\pm \,\mu} \az_\pm^\mu,\\
 \frac{v}{z} B_{\pm\,z}^{\ \mu} A_{\pm\,\mu} & \sim & -m \bz_{\pm\, \mu} \az_{\pm}^\mu,\\
\notag \sum_{a=\pm} B_{a\,\mu\nu}B_a^{\mu\nu} & \sim & 4 q^2 \log(Q\varepsilon)\left(q^2\Tz_+^\mu\Tz_{+\,\mu}+q^2{\tTz}_+^\mu{\tTz}_{+\,\mu}-\tlp m \Tz_+^\mu \az_{+\,\mu} \right) +\\
& & +8 q^2(\Tz_+^\mu\Tt_{+\,\mu}-{\tTz}_+^\mu{\tTt}_{+\,\mu})+\cdots
\end{eqnarray}
Where the dots refer to local terms in the sources, they will not be relevant because we can remove them with finite counterterms.

To this action we have to add some boundary counterterms to remove the divergences that appear as $\varepsilon\to 0$. As expected, the leading divergence $1/\varepsilon^2$ does not appear in the action of the two-form field. There are however additional logarithmic divergences. In order to completely cancel them we need more counterterms, of the form $H^2$, $F^2$, $(dX)^2$ and $XFB$. More explicitly, we have that the finite regularized action is
\begin{equation}
S_{reg}=S_{o.s}+S_1+S_2+S_{3\, +}+S_{3\, -}+S_4+S_{\rm finite}
\end{equation}
where
\begin{align}
&S_{1}=c_{1}\int d^4 x \log(\mu\varepsilon) \sqrt{-\gamma}  \sum_{a=\pm} H_{a\,\mu\nu\sigma} H_a^{\mu\nu\sigma},\\
&S_2=c_2\int d^4 x \log(\mu\varepsilon) \sqrt{-\gamma}  X_+ F_{V\, \mu\nu} B_+ ^{\mu\nu},\\
&S_{3,\pm}=c_{3,\pm}\int d^4 x  \log(\mu\varepsilon)\sqrt{-\gamma}  F_{\pm\,\mu\nu} F_\pm^{\mu\nu}.\\
&S_4=c_4\int d^4 x  \log(\mu\varepsilon)\sqrt{-\gamma} (D_\mu X)^\dagger (D^\mu X).
\end{align}
Notice that we can also have finite conterterms, corresponding to $S_{1}$, $S_2$, $S_{3\, \pm}$ and $S_4$ with no log factors. We will introduce the finite counterterms in $S_{\rm finite}$ and use them later on.

In the $\varepsilon\to 0$ limit, the counterterms become
\begin{eqnarray}
S_{1} & \sim&  -6 (q^2)^2 \Big(\Tz_{+\,\mu} \Tz_+^\mu+{\tTz}_{+\,\mu} {\tTz}_+^\mu\Big),\\
S_2 & \sim & -2 m q^2 \az_{+\,\mu} \Tz_+^\mu,\\
S_{3\,\pm} & \sim & -2 q^2  \az_{\pm\,\mu} \az_\pm^\mu,\\
S_4 &\sim & m^2 \az_-^\mu \az_{-\,\mu}.
\end{eqnarray}
One can cancel the quadratic and logarithmic divergences if
\begin{equation}
c_{1}=\frac{2g_B}{9}, \ \ c_2=-\frac{\lambda}{6}, \ \ c_{3\,\pm}=\frac{1}{4 g_5^2}, \ \  c_4=\frac{g_X^2}{2}.
\end{equation}
Now one can take the $\varepsilon\to 0$ limit, and use finite counterterms to remove the pieces that are local in the sources
\begin{multline}\label{eq:renaction}
S_{ren} = \int \frac{d^4 q}{(2\pi)^4} \left[-\frac{1}{g_5^2} \sum_{a=\pm} \az_{a\,\mu}\at_a^\mu+ \frac{8 g_B}{3} q^2(\Tz_+^\mu\Tt_{+\,\mu}-{\tTz}_+^\mu{\tTt}_{+\,\mu})\right.+\\\left.+ q^2\log\frac{Q^2}{\mu^2}\left(\frac{1}{4 g_5^2} \sum_{a=\pm}\frac{q_a^2}{q^2}\az_{a\,\mu} \az_a^\mu  -\frac{\lambda}{6}m\az_+^\mu\Tz_{+\,\mu}\right.\right.+\\\left.\left.+\frac{2g_B}{3} q^2 (\Tz_+^\mu\Tz_{+\,\mu}+{\tTz}_+^\mu{\tTz}_{+\,\mu})\right) \right].
\end{multline}


\subsection{Two-point functions and matching to QCD}


We now proceed to write the renormalized action in terms of general sources $v_\mu$ and $t_{\mu\nu}$. The transverse vector field is
\begin{equation}
\az_\mu=P_\mu^{\ \alpha} a_\alpha = \left(\delta_\mu^\alpha-\frac{q_\mu q^\alpha}{q^2} \right)a_\alpha.
\end{equation}
The transverse tensor is
\begin{equation}
t^T_{\mu\nu}= \frac{1}{2} P_{[\mu}^{\ \alpha} P_{\nu]}^{\ \beta} t_{\alpha\beta} = \frac{1}{2q^2} {F_+}_{\mu\nu}^{\ \ \alpha \beta} t_{\alpha\beta}.
\end{equation}
The longitudinal part of the tensor is
\begin{equation}
t^L_{\mu\nu}=\frac{1}{2}\left[(\delta_\mu^\alpha \delta_\nu^\beta-\delta_\mu^\beta \delta_\nu^\alpha)-P_{[\mu}^{\ \alpha} P_{\nu]}^{\ \beta}\right] t_{\alpha\beta} = -\frac{1}{2q^2} {F_-}_{\mu\nu}^{\ \ \alpha\beta} t_{\alpha\beta}
\end{equation}
Using the expansion
\begin{equation}
t_{\mu\nu}=i\epsilon_{\mu\nu\sigma\rho}q^\sigma\tTz^\rho+iq_{[\mu}\Tz_{\nu]},
\end{equation}
we indeed find
\begin{equation}
t^T_{\mu\nu}=i\epsilon_{\mu\nu\sigma\rho}q^\sigma\tTz^\rho,  \ \ t^L_{\mu\nu}=iq_{[\mu}\Tz_{\nu]}.
\end{equation}

We now assume that the subleading terms are proportional to the sources
\begin{eqnarray}
\notag \at_{+\,\mu} & = & G_{VV}(q^2) \az_{+\,\mu}+ G_{VT}(q^2) \Tz_{+\,\mu}, \\
\notag \Tt_{+\,\mu} & = & G_{TT}^-(q^2) \Tz_{+\,\mu}+G_{TV}(q^2) \az_{+\,\mu}, \\
{\tTt}_{+\,\mu} & = & G_{TT}^+(q^2) {\tTz}_{+\,\mu}. \label{eq:defG}
\end{eqnarray}
And using  \eqref{eq:ident}
\begin{eqnarray}
\notag \az_\mu \az^\mu & = & a_\mu P^\mu_{\ \alpha} P^{\alpha\nu} a_\nu = a_\mu P^{\mu\nu} a_\nu\\
\notag \Tz_\mu \Tz^\mu & = & \frac{1}{(q^2)^2} q^\alpha t^L_{\alpha\mu} q^\beta{t^L}_\beta^{\ \mu} = -\frac{1}{4 (q^2)^2} t_{\mu\nu} F_-^{\mu\nu;\alpha \beta} t_{\alpha\beta}\\
\notag \az_\mu \Tz^\mu & = & -\frac{i}{2 q^2} q_{[\mu} P_{\nu]}^\sigma a_{\sigma} {t^L}^{\mu\nu} = \frac{i}{2 q^2} a_\mu (\eta^{\mu\alpha}q^\beta- \eta^{\mu\beta}q^\alpha) t_{\alpha\beta}\\
\notag {\tTz}_{\mu} {\tTz}^{\mu} & = & \frac{1}{4(q^2)^2}\epsilon^{\mu\nu\sigma\rho} \epsilon^{\alpha\beta\gamma\delta}g_{\mu\alpha}q_\nu q_\beta\, {t^T}_{\sigma\rho} {t^T}_{\gamma\delta}=  - \frac{1}{4(q^2)^2} t_{\mu\nu} F_+^{\mu\nu;\alpha \beta} t_{\alpha\beta}.
\end{eqnarray}
Introducing these expressions in the renormalized action \eqref{eq:renaction} and deriving twice with respect to the sources, we find the following correlation functions
\begin{eqnarray}
\Pi_{VV}^{\mu\nu,\, ab}(q^2) & = & \delta^{ab}(q^\mu q^\nu-q^2 \eta^{\mu\nu}) \Pi_{VV}(q^2), \\
\Pi_{VT}^{\mu;\nu\rho,\, ab}(q^2) & = & i\delta^{ab}(\eta^{\mu\nu} q^\rho-\eta^{\mu \rho} q^\nu) \Pi_{VT}(q^2), \\
\Pi_{TT}^{\mu\nu;\alpha\beta,\, ab}(q^2) & = & \delta^{ab}\Pi_{TT}^+(q^2) F_+^{\mu\nu;\alpha\beta} +\delta^{ab}\Pi_{TT}^-(q^2) F_-^{\mu\nu;\alpha\beta}.
\end{eqnarray}
Where
\begin{equation}\label{eq:Pivv}
\Pi_{VV}(q^2) = -\frac{1}{2 g_5^2} \log\frac{Q^2}{\mu^2}+\frac{2}{g_5^2 q^2} G_{VV}(q^2),
\end{equation}
\begin{equation}\label{eq:Pivt}
\Pi_{VT}(q^2) =-\frac{\lambda}{12}m\log\frac{Q^2}{\mu^2} -\frac{1}{2 g_5^2 q^2} G_{VT}(q^2)+\frac{4}{3} g_B G_{TV}(q^2),
\end{equation}
\begin{equation}\label{eq:Pittp}
\Pi_{TT}^+(q^2) = -\frac{g_B}{3}\log\frac{Q^2}{\mu^2}+\frac{4}{3}\frac{g_B}{q^2} G_{TT}^+(q^2),
\end{equation}
\begin{equation}\label{eq:Pittm}
\Pi_{TT}^-(q^2) = -\frac{g_B}{3}\log\frac{Q^2}{\mu^2}-\frac{4}{3}\frac{g_B}{q^2} G_{TT}^-(q^2).
\end{equation}
Comparing with the expressions \eqref{eq:vvope}, \eqref{eq:ttope} and \eqref{eq:vtope} we get
\begin{equation}
\frac{1}{g_5^2}=\frac{N_c}{12\pi^2}, \ \  g_B=\frac{N_c}{16\pi^2}, \ \ \lambda = -\frac{3N_c}{4\pi^2}.
\end{equation}
Together with \eqref{eq:gX}, this fixes all the coupling constants of the bulk action.


\section{Matching to the massless theory}\label{ss:matching}


We have used the coefficients of the logarithmic divergences of the correlation functions to fix the parameters of the model. Notice that in the massless limit $m\to 0$ the logarithmic contribution to the vector-tensor correlator \eqref{eq:Pivt} vanish. In QCD, perturbative contributions vanish to all orders, so the only contributions left come from non-perturbative physics, this is clear in \eqref{eq:vtope}, where the leading term when the mass is zero is proportional to the condensate. Therefore, for massless QCD this is the term we have to match to fix the value of the parameter $\lambda$. Since the coefficient of this term is independent of the mass we should get the same value for $\lambda$, we will se that this is indeed the case, so the model passes this non-trivial consistency check.

In order to find the non-perturbative contributions to the OPE, we need to compute the functions `$G(q^2)$' that appear in \eqref{eq:defG} and plug them in the expressions for the correlators that we have found in the previous section.
The overall strategy will be to solve the relevant equations of motion and match the near boundary expansion of the solutions to the coefficients of the series defined in \eqref{eq:texp} and \eqref{eq:vecexp}.

As we mentioned before, there are four coupled equations describing the negative parity mesons. Plugging \eqref{eq:bexp} in equations \eqref{eq:vecb}, \eqref{eq:vecv}, \eqref{eq:bzp}, \eqref{eq:tensorp} and \eqref{eq:tensorm} and keeping only the negative parity modes ($V, T^{\nu}_+, \overline{T}_-^{\nu}, B_+^{z \nu}$), we have the equations:
\begin{eqnarray}
&& \left[\partial_z^2-\frac{f(z)}{g(z)}\partial_z-(C_1(z)-q^2)\right]B_+^{\nu z}=-\frac{\lambda}{8 g_B}\left[C_2(z)V'^{\nu}+\frac{v'(z)q^2}{g(z)}V^{\nu}\right],\\
 \label{eq:bbz}
 &&\partial_z^2V^{\nu}-\frac{f(z)}{g(z)}\partial_zV^{\nu}+q^2V^{\nu}=\frac{\lambda g_5^2v'(z)}{g(z)}B_+^{\nu z},\\
\label{eq:vv}
&&\left[ z\partial_zz\partial_z-1+z^2q^2\right]T^{\nu }_+=-\frac{\lambda}{8 g_B}\left[v(z)V^{\nu}-\frac{z^2v'(z)}{g(z)}\partial_zV^{\nu}\right]+
\left(z^2\frac{f(z)}{g(z)}+z\right)B_+^{\nu z},
\label{eq:tensorp2}\\
&&\left[ z\partial_zz\partial_z-1+z^2q^2\right]\overline{T}_-^{\nu}=-\frac{\lambda}{8 g_B} zv'(z)V^{\nu}
\label{eq:tensorm2}
\end{eqnarray}
With the aditional constraint:
\begin{eqnarray}
\partial_zB^{\nu z}_+-\frac{f(z)}{g(z)}B^{\nu z}_++iq^2T^{\nu}_+=\frac{\lambda}{8g_B}\frac{v'(z)}{g(z)}\partial_zV^{\nu}.
\label{eq:constrain}
\end{eqnarray}
The mixing term proportional to $\lambda v(z)$ in the equations of motion is a small perturbation for small values of $z$, as $v(z)$ falls off towards the boundary. As the large $Q$ behavior of spatial correlators with $Q^2=-q^2$ is dominated by the small $z$ behavior of the solution, we can determine the short distance behavior of correlation functions analytically by treating $\lambda$ as a small parameter and solving the the equations of motion  perturbatively in $\lambda$. However, we don't need to solve all the four equations. First, the constraint above implies that $B^{\nu z}$ is not independent, and the relations \eqref{eqcond} imply that close to the boundary, $T_+^{\nu}$ is not independent of $\overline{T}^{\nu}_-$. For convenience, we will focus on equations \eqref{eq:vv} and \eqref{eq:tensorm2}. Dropping all terms of order $\lambda^2$ in the equations above, and taking the ansatz:
\begin{eqnarray}
&& V^{\nu}=V_0(x)v^{\nu}+V_{\lambda}(x)b^{\nu}\\
&& \overline{T}^{\nu}_-=T_0(x)\overline{t}_-^{\nu}+T_{\lambda}(x)v^{\nu}
\end{eqnarray}
Our problem is reduced to solving the equations below
\begin{eqnarray}
&&\left(\partial_x^2-\frac{1}{x}\partial_x+q^2\right)V_0(x)=0,
\label{eq:vechom}\\
&&\left(\partial_x^2-\frac{1}{x}\partial_x+1\right)V_{\lambda}(x)=\lambda(\alpha_1 +\alpha_2x^2)B_0(x),
\label{eq:vecl}\\
&&\left[ x^2\partial^2_x+\partial_x-1-x^2\right]\overline{T}_0(x)=0,
\label{eq:tensorhom}\\
&&\left[ x\partial^2_x+\partial_x-\frac{1+x^2}{x}\right]\overline{T}_{\lambda}(x)=-\lambda (\Gamma_1x+\Gamma_2x^3)V_0(x).
\label{eq:tensorl}
\end{eqnarray}
Where $\alpha_1=\frac{g_5^2m}{Q^2}$, $\alpha_2=\frac{3g_5^2\sigma}{Q^4}$, $\Gamma_1=\frac{m}{8g_BQ}$ and $\Gamma_2=\frac{3\sigma}{8g_BQ^3}$.
$V_0$, $\overline{T}_0$ are the homogeneous solutions and $V_{\lambda}$, $\overline{T}_{\lambda}$ the perturbative corrections
of order $\lambda$. $B_0$ satisfies the same equation as $V_0$.
Focusing first on the vector mode, $V_0$ and $B_0$ have well known solutions in terms of Bessel functions:

\begin{eqnarray}
\label{eq:vvv1}
&&V_0(x)=x(aI_1(x)+K_1(x)),\\
\label{eq:bbb}
&&B_{0}(x)=x(bI_1(x)+K_1(x)),
\end{eqnarray}
Where $a$ and $b$  are constants fixed after imposing the IR boundary condition.
For $V_0(x)$, following previous work we are going to choose a Neumann boundary condition at $x_m$, $\partial_xV_0(x)\big|_{x_m}=0$ which allows us to set $a=\frac{K_0(x_m)}{I_0(x_m)}$. To choose the appropriate boundary condition for $B_0$ we note that a Dirichlet boundary condition imposed on $B_0(x)$, which sets $b=-\frac{K_1(x_m)}{I_1(x_m)}$, implies, by use of the constrain \eqref{eq:constrain} to leading order, a Neumann boundary condition for the leading tensor mode, but  a choice of Neumann boundary condition for the former is not consistent.
As $B_0$ and $V_0$ describe completely independent modes, this already captures the most general solution.\\

To compute $V_{\lambda}$ we will use a Green's function method. Note that we can write a solution for equation \eqref{eq:vecl} of the form:
\begin{equation}
 \label{eq:1orderl2}
V_{\lambda}(x)=\lambda\int^{x_m}_{x_{\varepsilon}} dx'[\alpha_1+\alpha_2x'^2]B_0(x')\frac{G_V(x,x')}{x'}.
\end{equation}
Provided $G_V$ satisfies the equation:
\begin{equation}
\left(\partial^2_x-\frac{1}{x}\partial_x-1\right)G_V(x,x')=x\delta(x-x').
\label{eq: green}
\end{equation}
With boundary conditions $G_V(x_{\varepsilon},x')=G_V'(x,x_m)=0$.
We solve the equation above in the two regions $x>x'$ and $x<x'$ and match the two solutions at $x=x'$. It is not hard to show that $G_V(x,'x)$ can be written as:
\begin{equation}
G_V (x,x')=\frac{xx'}{AD-BC}[AI_1(x_>)+BK_1(x_>)][CI_1(x_<)+DK_1(x_<)].
\label{eq:greens}
\end{equation}
Where $x_{<,>}=\{min,max\}(x,x')$ is book keeping notation to specify the two branches of the Green's function.
The coefficients are $A=-K_0(x_m)$; $B=I_0(x_m)$;$C=K_1(x_{\varepsilon})$;$D=-I_1(x_{\varepsilon})$. Taking the limit $x_{\varepsilon} \rightarrow 0$ we can set above $D=0$ and $C=1$.
Replacing back in \eqref{eq:1orderl2} we have:
\begin{eqnarray}
V_{\lambda}(x)=-\lambda\int^x_{0}dx'[\alpha_1x'+\alpha_2x'^3]K_1(x')I_1(x')-\lambda\frac{x^2}{2}\int^{x_m}_{x}dx'[\alpha_1x'+\alpha_2x'^3]K^2_1(x').
\label{eq:bc1}
\end{eqnarray}
Where we used:
\begin{eqnarray}
&&\frac{x(CI_1(x)+DK_1(x))}{AD-BC}=-xI_1(x)\simeq-\frac{x^2}{2},\\
&&\frac{x(AI_1(x)+BK_1(x))}{AD-BC}\simeq -1,
\end{eqnarray}
and
\begin{equation}
B_0\simeq xK_1(x).
\end{equation}
We also have ignored all the terms proportional to $\frac{A}{B}\sim e^{-2x_m}$ since in the limit of large momentum these terms vanish quickly. Physically this means that as the momentum increases what happens in the  IR region becomes less important, as expected. In fact, these solutions near the boundary and for large momentum become oblivious of the IR boundary conditions, since both Neumann and Dirichlet boundary conditions will enforce factors that fall off exponentially.
Moroever, we will ignore the contribution of the first integral, that is negligible since these contributions vanish too quickly near the boundary.

We then have:
\begin{eqnarray}
V_{\lambda}(x)=-\lambda\frac{x^2}{2}\int^{x_m}_{x}dx'(\alpha_1x'+\alpha_2x'^3)K_1(x')K_1(x')
\simeq \lambda\alpha_1\frac{x^2}{2}\left(\frac{1}{2}+\log x\right)-\lambda\frac{x^2}{3}\alpha_2.
\label{eq:solvl}
\end{eqnarray}
The near boundary solution for $V^{\nu}$ is:
\begin{equation}
 V^{\nu}(x)=V_0(x)v^{\nu}+V_{\lambda}b^{\nu}=\left(1-\frac{x^2}{4}+\frac{x^2}{2}\log x\right)v^{\nu}+\left(\lambda\alpha_1\frac{x^2}{2}\left(\frac{1}{2}+\log x\right)-\lambda\frac{x^2}{3}\alpha_2\right)b^{\nu}.
\end{equation}
Matching the solution above to the expansion defined in \eqref{eq:vecexp}, and using \eqref{eqcond}, we find:
\begin{equation}
 A^{(2)}_{+ \mu}\equiv V^{(2)}_{\mu}=-\frac{Q^2}{4}V^{(0)}-\lambda Q^4 \left(\frac{\alpha_1}{4}-\frac{\alpha_2}{3}\right)T^{(0)}_+.
\end{equation}
Therefore:
\begin{equation}
 G^+_{VV}(Q^2)=-\frac{Q^2}{4},\,\,\,\,\,G^{+}_{VT}(Q^2)=-\lambda Q^4 \left(\frac{\alpha_1}{4}-\frac{\alpha_2}{3}\right).
\end{equation}
Following similar steps, we can now compute near boundary solutions for equations \eqref{eq:tensorhom} and \eqref{eq:tensorl}.
The homogeneous equation has a well known solution of the form:
\begin{equation}
 \overline{T}_-(x)=cI_1(x)+K_1(x)\simeq K_1(x).
\end{equation}
For the second equation, again, we can write a solution with a Green's function:
\begin{equation}
 \overline{T}_{\lambda}(x)=-\lambda\int^{x_m}_{x_{\epsilon}}\,dx'(\Gamma_1x'+\Gamma_2x'^3)K_1(x')G_T(x,x'),
\end{equation}
where analogously to the previous calculation, $G_T(x,x')$ satisfies:
\begin{equation}
 \left[x\partial^2_x+\partial_x-\frac{1+x}{x}^2\right]G_T(x,x')=\delta(x-x').
\end{equation}
It can be shown that:
\begin{equation}
 G_T(x,x')\simeq-K_1(x_>)I_1(x_<),
\end{equation}
in the limit where $x_{\epsilon} \rightarrow 0$ and $x_m \rightarrow \infty$. Finally, we find that the solution for the tensor field is:
\begin{eqnarray}
\overline{T}_{\lambda}(x)=\lambda\frac{x}{2}\int^{\infty}_x\,dx'[(\Gamma_1x'+\Gamma_2x'^3)K_1(x')]K_1(x')=\lambda\left(\frac{\Gamma_2}{3}-\frac{\Gamma_1}{4}\right)x,
\end{eqnarray}
therefore,
\begin{equation}
\label{eq:tm}
 \overline{T}^{(2) \mu}_-=-\frac{Q^2}{4}\overline{T}^{(0)\mu}_-+Q\left(\frac{\Gamma_2}{3}-\frac{\Gamma_1}{4}\right)V^{(0) \mu}.
\end{equation}
However, we are really after $T^{(2) \mu}=G_{TT}(Q^2)T^{(0) \mu}+G^+_{TV}(Q^2)V^{(0) \mu}$. To compute the latter, we use the relations we have found previously:
\begin{eqnarray}
 &&\overline{T}^{(2) \mu}_-=- T_+^{(2) \mu}+\frac{1}{2}\left(q^2T^{(0) \mu}_++\frac{\lambda}{8g_B}m V^{(0) \mu}\right)\\
&&\overline{T}^{(0)}_{-}= T^{(0) \mu}_+.
\end{eqnarray}
Solving for $T^{(2) \mu}_+$ we get:
\begin{equation}
T^{(2) \mu}_+ =-\overline{T}^{(2) \mu}_-+\frac{1}{2}\left(q^2T^{(0) \mu}_++\frac{\lambda}{8g_B}m V^{(0) \mu}\right)=-\frac{Q^2}{4}T^{(0) \mu}_++\left[\frac{5\lambda m}{32G_B}-\frac{\sigma\lambda}{32 g_BQ^2}\right]V^{(0) \mu}.
\end{equation}
So the result is,
\begin{eqnarray}
 G^+_{TT}(Q^2)=-\frac{Q^2}{4},\,\,\,\,\,G_{TV}^+(Q^2)=\frac{5\lambda m}{32}-\frac{\sigma\lambda}{8 g_BQ^2}.
\end{eqnarray}

We can now compare with the OPE of vector and tensor correlators (\ref{eq:Pivv}-\ref{eq:Pittm}).
Setting the mass to zero, the only nonzero contributions are
\begin{eqnarray}
G_{VV}=-\frac{Q^2}{4}, & G_{VT}=\lambda g_5^2 \sigma, \\
G_{TT}^\pm = -\frac{Q^2}{4} & G_{TV} = -\frac{\lambda\sigma}{8g_BQ^2}.
\end{eqnarray}
The contributions $G_{VV}$ and $G_{TT}$ give contact terms that can be removed using counterterms in the regularized action. The only non-perturbative contributions to this order are
\begin{equation}
\Pi_{VT}(q^2)=\left(\frac{1}{2}-\frac{2}{3}\right)\frac{\lambda \sigma}{Q^2} = -\frac{\lambda}{6 g_X^2}\frac{\langle \overline{\psi}\psi\rangle}{Q^2} =-\frac{\langle \overline{\psi}\psi\rangle}{Q^2}.
\end{equation}
Comparing with the OPE in QCD \eqref{eq:vtope}, we see this term has the right coefficient, including the sign. The results we have obtained are insensitive to the details of the IR, so they should be valid for any models that are asymptotically AdS space. However, the value of the condensate itself and other quantities like the meson spectrum will be sensitive to IR physics. In the next section we will study how the inclusion of the new terms in the action affect to some of these quantities.


\section{Meson spectrum}\label{ss:spectrum}


So far we have discussed the UV physics of our model, focusing in the matching with the OPE of correlators in QCD. We will now comment on some of the IR physics, in particular the meson spectrum. In our analysis we have seen that the two-form field splits in a transverse part and a longitudinal part, that mixes with the vector fields. We can summarize the correspondence between the fields and meson states in the following table:
$$
\begin{array}{ccc}
B_{\mu\nu} & {\rm mixes} & J^{PC} \; {\rm mesons} \\ \hline
{\rm transverse} & - & 1^{+-} \\
{\rm longitudinal} & V_\mu & 1^{--} \\
\end{array}
$$
The lightest isospin triplet states that can be found in the Particle Data Group (PDG) review \cite{Nakamura:2010zzi}, are
$$
\begin{array}{ccc}
{\rm meson} & J^{PC} & {\rm mass \; (MeV)} \\ \hline
b_1(1235) & 1^{+-} & \sim 1229.5\pm 3.2 \\
\rho(770)  & 1^{--} & \sim 775.49\pm 0.34 \\
a_1(1260) & 1^{++} & \sim 1230\pm 40 \\
\end{array}
$$
Notice that according to the PDG estimate, the $b_1$ and $a_1$ mesons are almost degenerate, although the error in the estimate of the $a_1$ mass is very large. Other estimates give a mass to the $a_1$ $\sim 1255$ MeV, with somewhat smaller errors \cite{Alekseev:2009xt}.

In the holographic model the $b_1$ state is obtained from the transverse components of the $B$ field, that are decoupled from the rest of the fields.
The degeneracy between the $\rho$ and the $b_1$ is broken in our model, thanks to the interaction term proportional to $\lambda$ in ref. \eqref{eq:action1}. Had we not considered this term, the spectrum would be degenerate, as has been observed in ref. \cite{Cappiello:2010tu}. So we should include this cubic interaction term both from the perspective of the large momentum OPE and from the properties of the meson spectrum.

We follow a similar procedure as in ref. \cite{Erlich:2005qh} to compute numerically the lowest masses of the vector meson spectrum. We must specify suitable boundary conditions for the fields at the IR radial cutoff $z=z_m$ (Neumann or Dirichlet) and at the boundary $z=0$ (normalizability). Solutions do not exist for any value of the four-momentum $q^2$, but only for a discrete set of values, which correspond to the masses of mesons in the holographic dual $m_n^2=q^2$. We have checked that our results for the meson spectrum and the pion decay constant $f_\pi$ coincide with those of ref. \cite{Erlich:2005qh} when we set the coupling $\lambda=0$.

We start with the spectrum of $1^{+-}$ mesons, dual to the field components $\{{P}^\mu_\alpha {P}^\nu_\beta B_+^{\alpha\beta}\}$ and $\{ B_-^{\mu z}, (\delta^\mu_\alpha-{P}^\mu_\alpha) B_-^{\alpha \nu}\}$. Notice that we can solve first for $B_-^{\mu z}$ in \eqref{eq:bzm} and then use \eqref{eq:tensorm} to solve for $(\delta^\mu_\alpha-{P}^\mu_\alpha) B_-^{\alpha \nu}$. As we have explained \eqref{eq:tensorp} is equivalent to \eqref{eq:bzm}, so for the purpose of finding the masses it is enough to focus on \eqref{eq:bzm}. Close to the boundary, a normalizable solution has the asymptotic expansion \eqref{eq:vecexp} with $\bz_-^{\mu}=0$. At the cutoff we impose Neumann boundary conditions, since for Dirichlet boundary conditions there is a normalizable solution at $q^2=0$, which would be dual to a massless vector meson. Normalizable solutions are Bessel functions $B_-^{\mu z}= b^\mu z J_1(|q| z)$ and the Neumann boundary condition is satisfied for values of the momentum such that $J_0(|q| z_m)=0$. Then, the mass of the lowest mode is
\begin{equation}
m_{b_1} z_m \simeq 2.405.
\end{equation}
Notice that this value is independent of the quark mass and condensate. The remaining modes do depend on them and we have to solve numerically the equations.

We will first solve for modes dual to pseudoscalar mesons, whose lowest mode corresponds to the pion. We need to solve the set of equations \cite{Erlich:2005qh}
\begin{align}
&\varphi''-\frac{1}{z}\varphi'+g_5^2 g_X^2 \frac{v}{z^3} (\pi-\varphi)=0,\\
&-q^2\varphi'+\frac{g_5^2 g_X^2 v^2}{z^2} \pi'=0.
\end{align}
For this, we first derive a single second order equation by solving algebraically for $\pi$ in the first equation, plugging the result in the second equation and defining $\phi=\varphi'$. Then, using $g_X^2 g_5^2=3$ and defining $h(z)=3 v(z)^2/z^3$, we obtain
\begin{equation}
\phi''(z)+\frac{h'}{h}\phi(z)-\left(\left(\frac{h'}{h}\right)^2-\frac{h''}{h}+z h-q^2 \right) \phi(z)=0.
\end{equation}
Normalizable solutions at the boundary behave as $\phi(z)\sim z$ and we impose a Dirichlet boundary condition at the cutoff for the field $\phi$. Then, for given values of $m z_m$ and $\sigma z_m^3$ we find the lowest value of $q_1^2 z_m^2=m_\pi^2 z_m^2$ such that a solution satisfying the boundary conditions exists. We can then use the physical value of the pion mass $m_\pi=139.6$ MeV to fix the scale $z_m$.

The spectrum of axial vector mesons $1^{++}$ can be found by solving equation \eqref{eq:em2}. From \eqref{eq:vecexp} a normalizable solution $\az_-^\mu=0$ vanishes at the boundary. At the cutoff, we impose Neumann boundary conditions. Finally, the spectrum of vector mesons can be computed from the system of coupled equations \eqref{eq:vecv} and \eqref{eq:bzp}, with conditions $\az_+^\mu=0$, $\bz_+^\mu =0$ in the expansions at the boundary \eqref{eq:vecexp}. Regarding the boundary conditions at the cutoff, we must be careful since equations \eqref{eq:bzp} and \eqref{eq:tensorp} have an additional singular point at $z_*$ such that $g(z_*)=1$. We are then constrained to values of the quark mass and the condensate such that $z_*>1$ or to impose suitable boundary conditions at the singular point. A quick analysis shows that the two possible behaviors of solutions close to the singular point are $\sim (z_*-z)^{1/2}$ and $\sim 1$ for $V^\mu$ and $\sim (z-z_*)^{(1\pm\sqrt{13})/4}$ for $B_+^{\mu z}$. We can then make the solution regular by imposing a Neumann boundary condition for $V^\mu$ and a Dirichlet boundary condition for $B_+^{\mu z}$.

For the values of the mass and the condensate we have explored $0.0001\leq m z_m \leq 0.1$, $0.0125\leq\sigma z_m^3\leq 0.5$ we do not find a realistic spectrum of mesons, the lightest vector meson $1^{--}$ is always heavier than both parity even mesons $1^{++}$, $1^{+-}$. For larger values of the mass we can understand this as a consequence of the singularity at $z=z_*$. The curvature of $AdS$ makes the classical problem of finding normalizable modes effectively as the quantum mechanical problem of finding the energy spectrum of a particle in a box, with one of the walls at the cutoff. For the $1^{--}$ modes we are forced to impose boundary conditions at the singularity $z_*<z_m$, so the ``box'' is smaller and the spectrum is lifted to higher values. This could be a problem of how infrared effects are implemented in this particular model, maybe different constructions like the soft wall could avoid this issue.

There is a way to find a more realistic meson spectrum, with the parity odd vector meson below the other modes. Instead of introducing the cutoff, we can impose boundary conditions for the vector mesons at the singularity even when it sits at a radial position beyond the cutoff $z_*>z_m$. For large enough values, the vector mesons become lighter and the spectrum can be tuned to realistic values, for instance for $mz_m=0.0005$, $\sigma z_m^3=0.05375$ we find that $m_\rho \simeq 753.95$ MeV, $m_{a_1}\simeq 1238.24$ MeV and $m_{b_1}\simeq 1237.87$ MeV. Although this would fix the meson spectrum, there are other quantities that are important to determine whether the model is phenomenologically viable. One such quantity is the pion decay constant, $f_\pi$, that in QCD is approximately $f_\pi\simeq 91.92$ MeV. In the holographic model it is given by the formula \cite{Erlich:2005qh}
\begin{equation}
f_\pi^2=-\frac{1}{g_5^2}\left.\frac{\partial_z A(z)}{z}\right|_{z=\varepsilon},
\end{equation}
where $A(z)$ is a solution to \eqref{eq:em2} satisfying $A(\varepsilon)=1$, $A'(z_m)=0$. With the parameters that give a realistic meson spectrum, the value of the pion decay constant is quite low $f_\pi\simeq 4.07$ MeV.


\section{Conclusions}\label{ss:conclusions}


We have carried out, for the first time, a complete treatment of the hard-wall model including all fields dual to operators of free field theory dimension 3. We followed the standard procedure of fixing bulk parameters by matching the short distance behavior of correlation functions to perturbative QCD. Reassuringly, the structure of the correlators we obtained from our holographic model precisely matched the expressions in perturbative QCD, so this program can be carried out consistently. With this matching in hand, we calculated physical properties of mesons which, unfortunately, no longer match QCD. While this result casts into doubt whether the simple hard wall model can serve as a good stand-in for QCD, one may hope that an improved IR model could potentially lead to a better spectrum. As our analysis of the short-distance behavior of correlation functions only relies on the UV asymptotics of the geometry, the action we derived (including the numerical values of the coupling constants) should serve as the starting point for any such exploration of complete (in the sense of including all dimension 3 field theory operators) holographic bottom-up models with alternative IR boundary conditions. As we discussed in section \ref{ss:equations}, it is possible to modify the bulk action of the two-form field by adding a kinetic term, giving a one-parameter family of theories with the desired self-duality condition and asymptotic behavior. Since this will modify the boundary action, in principle the value of the bulk couplings will be shifted when the matching to QCD is done. It is possible then, that by changing this parameter, a more realistic spectrum can be found.

Let us point out some differences between our approach and what one expects in a top-down models like Sakai-Sugimoto \cite{Sakai:2004cn,Sakai:2005yt}, based on a string theory construction. The matter content of the model is such that it coincides with large-$N_c$ QCD at low energies in some region of parameter space where the UV theory is weakly coupled. In particular, $1^{+-}$ mesons should be part of the spectrum. However, in the holographic description where supergravity is valid such modes are missing. This should not come as a surprise: since the tensor operator is not a BPS protected operator, its conformal dimension can receive large corrections of order $\sim \lambda^{1/4}$, where $\lambda$ is the 't Hooft coupling. In the holographic description this means that the tensor operator is dual to a field with a mass of order of the string scale, and therefore beyond the supergravity approximation. Since corrections to non-BPS operators are very large, it is even possible that the lowest $1^{+-}$ meson is not described by a field dual to the tensor operator we have considered in our model, but to a different operator with the same quantum numbers but larger conformal dimension in the free theory. This indeed seems to be the case in the Sakai-Sugimoto model, where the $1^{+-}$ mode is described by some components of a symmetric field in the bulk \cite{Imoto:2010ef}. Clearly, in this case we do not expect that the OPE of the model will match with that of QCD, so in some sense the approach of refs. \cite{Cappiello:2010tu,Domokos:2011dn} is closer to the top-down model. However, if the dimension of the tensor operator is chosen to be larger than $3$, it is more difficult to argue that the effective theory description in the bulk stays valid anymore.

We have studied the extension of the model that takes into account $1^{+-}$ mesons, like $b$ and $\omega$. In principle the model can be further extended to include other modes in the QCD spectrum that have been observed experimentally. A mode that is somewhat heavier, but not that much, than vector and axial vector modes is the $\pi_1(1400)$ meson, with $J^{PC}=1^{-+}$ and a mass $m_{\pi_1}\sim 1354\pm 25$ MeV \cite{Nakamura:2010zzi}. A peculiarity of this mode is that it cannot be predicted within the valence quark model, or in other words a simple quark bilinear operator would not create this kind of mode. An operator with the right quantum numbers would involve also a gluon field $\overline{\psi} F_{ij} \gamma_5 \psi$. Then, in order to include mesons with the quantum numbers of $\pi_1$, we would have to introduce a field dual to the dimension-five operators $\overline{\psi} F_{\mu\nu} \gamma_5 \psi$, and $\overline{\psi} F_{\mu\nu} \psi$. The obvious candidate is again a complex two-form field, with bulk mass $m^2\ell^2=9$ and no Chern-Simons action, since there is no self-duality constraint for these operators.


\section*{Acknowledgments}


We thank Gary Howell and Dam Son for collaboration during early stages of this work. We also thank Oscar Cata, Tuhin Roy and Shigeki Sugimoto for useful discussions. This work was supported in part by the U.S. Department of Energy under Grant No. DE-FG02-96ER40956.

\bibliography{bibmesons}
\end{document}